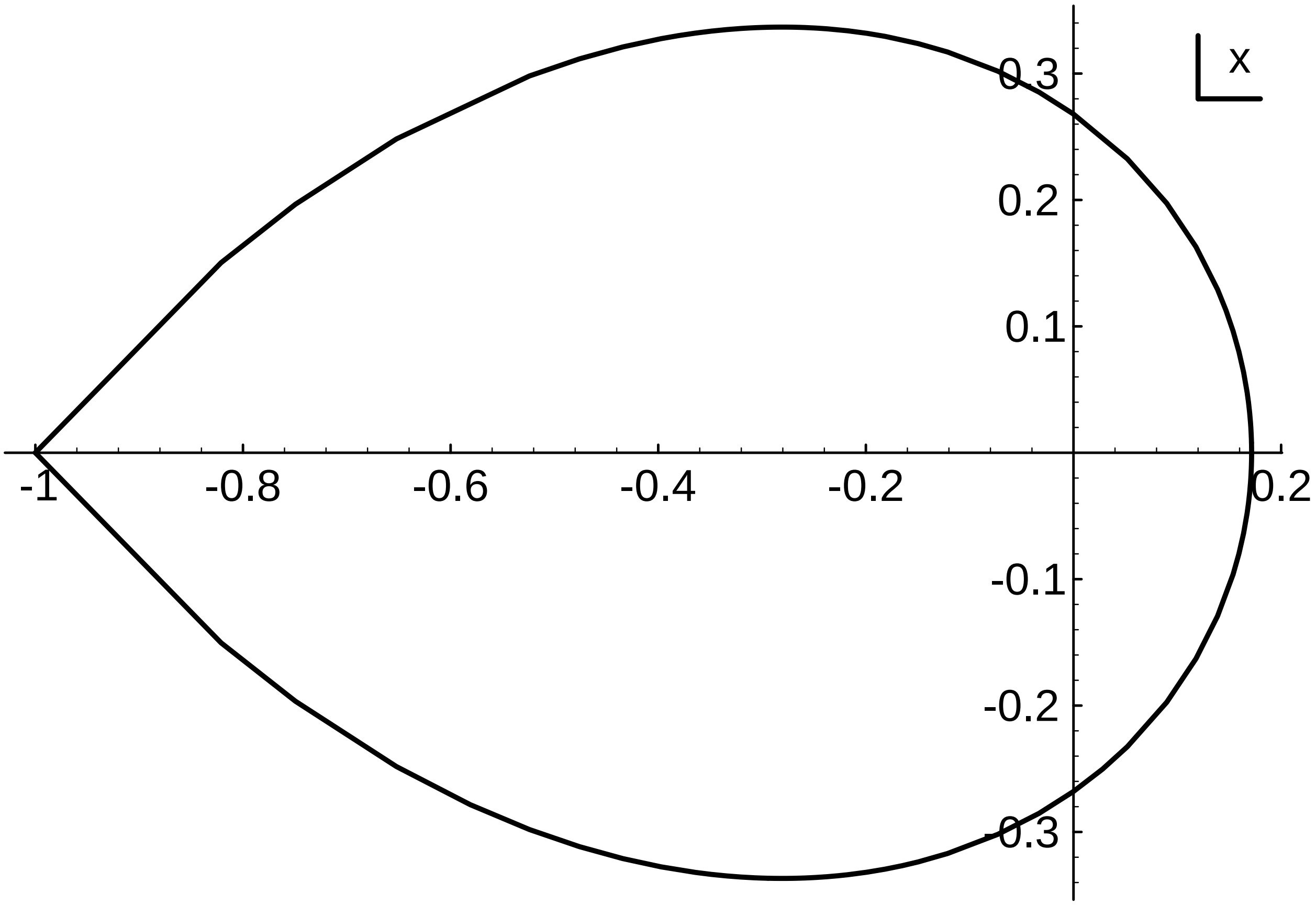

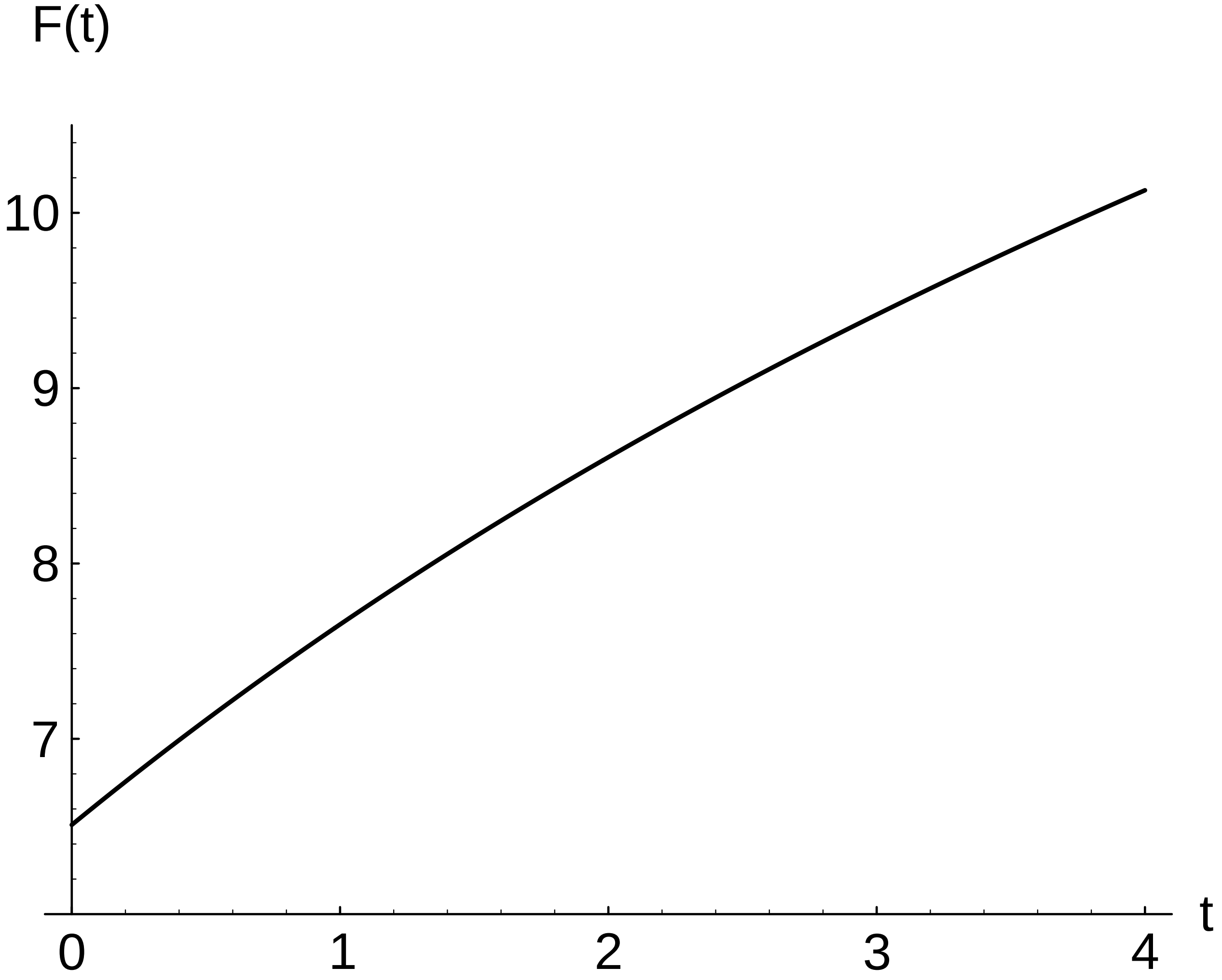

# Bethe ansatz solution for crossover scaling functions of the asymmetric XXZ chain and the KPZ-type growth model

Doochul Kim
*Department of Physics, University of Washington, Seattle, WA 98195 USA and Department of Physics, Seoul National University, Seoul 151-742, Korea**

A perturbative method is developed to calculate the finite size corrections of the low lying energies of the asymmetric XXZ hamiltonian near the stochastic line. The crossover from isotropic to anisotropic, Kardar-Parisi-Zhang (KPZ) scaling of the mass gaps is determined in terms of universal crossover scaling functions. At the stochastic line, the asymmetric XXZ hamiltonian describes the time evolution of the single-step or body-centered solid-on-solid growth model in one dimension. The mass gaps of the growth model are found as a function of the growth rate and the substrate slope. Higher order corrections to the growth model mass gaps are also calculated to obtain the first terms of the KPZ to Edwards-Wilkinson crossover scaling function in the large argument expansion in the zero slope sector.

## I. INTRODUCTION

Understanding and classifying dynamic scalings of self–affine growing surfaces has been an area of much interest [1]. One of the universality classes which attracted much attention is that of Kardar-Parisi-Zhang (KPZ) [2]. The KPZ equation is a nonlinear stochastic equation of motion for the height of growing surfaces whose local growth velocity depends nonlinearly on the local slope, and is equivalent to the noisy Burgers equation. When the nonlinear term is absent, the equation reduces to the diffusion equation termed as the Edwards-Wilkinson (EW) type. There are a number of microscopic lattice models which are believed to be described by the KPZ equation in continuum limit, e.g. the Eden model, the ballistic deposition model and the restricted solid on solid model [1]. However, identification of their universality class has been largely based on qualitative arguments and numerical evidences. An exception is the single–step, or the body-centered solid-on-solid model introduced by Plischke *et al.* [3] in one-space one-time dimension to which this work is addressed. In this model, the height of each site must differ by $\pm 1$ from its two neighbors and obey the following dynamic growth rule: Select one of the sites at random; if it is a local valley, i.e. if both of its neighboring sites are higher, its height is increased by 2 with the deposition probability $p_d$; if it is a local hill, its height is decreased by 2 with the evaporation probability $p_e$; otherwise, the height does not change. Due to the solid-on-solid condition, height configurations of this model have a natural representation in terms of the step variable $\{\sigma_i^z\}$ ($\sigma_i^z = \pm 1$). If one interpret the sites with $\sigma_i = 1$ as those occupied by a lattice gas particle, the model is that of a driven diffusive lattice gas [4]. The time evolution of the probability distribution for the state of the model is generated by the $\widetilde{\Delta} = 1$ line, or the stochastic line, of the following asymmetric XXZ hamiltonian [5,6];

$$\mathcal{H} = \sum_{i=1}^{N} \{ \frac{\widetilde{\Delta}}{4}(1 - \sigma_i^z \sigma_{i+1}^z) - \frac{1+s}{2}\sigma_i^+ \sigma_{i+1}^- - \frac{1-s}{2}\sigma_i^- \sigma_{i+1}^+ \} \tag{1}$$

where $N$ is the number of sites, $\sigma_i^{\pm}$ are the spin 1/2 raising and lowering operators, $\sigma_{N+1}^{\alpha} = \sigma_1^{\alpha}$ and $s = (p_d - p_e)/(p_d + p_e)$ is the growth rate. $m \equiv \sum_{i=1}^{N} \sigma_i^z / N$ is the mean slope of the substrate which is conserved in this model.

The hamiltonian with $\widetilde{\Delta}$ as a parameter has other physical applications as well. For example, it describes endpoints of facet–ridges in equilibrium crystal shapes and persistent current transitions in mesoscopic metallic rings [6–8]. The hamiltonian is an anisotropic limit of the transfer matrix of the general asymmetric six vertex model. When $s = 1$, the latter becomes the five-vertex model which describes, among others, an interacting domain wall system [9].

When $\widetilde{\Delta} < 1$, the hamiltonian is critical and possesses the conformal invariance [10]. Thus the mass gaps, which are the differences of low lying energy eigenvalues from the ground state energy, scale as $1/N$, the proportionality constant being the conformal dimension multiplied by a non-universal anisotropy factor. At the stochastic line, the ground state of the hamiltonian becomes trivial with zero energy in each sector, a subspace of $\mathcal{H}$ with fixed value of $m$. The mass gaps in this case are expected to scale as $1/N^{3/2}$, since the dynamic critical exponent $z$ of the KPZ universality class are known to be 3/2 in one dimension [11]. That this is indeed so has been demonstrated by Gwa and Spohn [5,12] from an analysis of the Bethe ansatz equation for the special case of $s = 1$, and $m = 0$. Also, recently Neergaard and den Nijs [6] have analyzed numerically obtained mass gaps to investigate the crossover behaviors from the KPZ line $0 < s \leq 1$ to the $s = 0$ EW point where the mass gaps scale as $1/N^2$. The bulk properties of the asymmetric six-vertex model at and near $\widetilde{\Delta} = 1$ are studied by Bukman and Shore [13,7] in connection with its application to the equilibrium crystal shape. Correlations in the asymmetric XXZ chain with reflecting boundary conditions which has finite energy gaps in the bulk is studied by Sandow and Schütz [14] and by Henkel and Schütz [15] in the context of driven diffusive lattice gas system. Also, the continuum limit of Eq. (1) at $\widetilde{\Delta} = 1$ is derived by Fogedby et al. [16] from which the dispersion relation $\omega \propto k^{3/2}$ is obtained for $s > 0$ in the quasi-classical limit. However, it is of interest to extend our exact knowledge of this model which can play a central role in the non-equilibrium growth problems.

In this work, we analyze the finite size corrections of the energy $E_N(\widetilde{\Delta}, s, m)$ of Eq. (1) for low lying levels using the Bethe ansatz equation and develop a systematic expansion of $E_N$ in powers of $1/N^{1/2}$ for general $s$ and $m$ and for small $1 - \widetilde{\Delta}$. This method enables one to obtain general expressions for mass gaps associated with various energy levels generalizing the work of Gwa and Spohn considerably. In the next section, we review the Bethe ansatz equation and put it in the form suitable for analysis for small $1 - \widetilde{\Delta}$. In Section III, finite sums are expressed as an infinite series in $1/N^{1/2}$ assuming certain analyticity property of the phase function introduced in Section II. The sum formula derived in Section III is then used in Section IV to determine the phase function self-consistently and to express the energy of Eq. (1) in the form of a perturbation series in $1/N^{1/2}$. In the next section, we solve explicitly the general expressions of Section IV to obtain leading order behaviors of $E_N$. To the leading order, mass gaps are then obtained in terms of universal scaling functions which describe the crossover from the isotropic ($\widetilde{\Delta} < 1$, $z$=1) to anisotropic ($\widetilde{\Delta}$=1, $z$=3/2) scaling behaviors. The scaling variable is found to be $s^{-1}(1 - m^2)^{1/2}(1 - \widetilde{\Delta})N^{3/2}$ up to an arbitrary numerical factor. In particular, the real part of the mass gaps are proportional to $s(1 - m^2)^{1/2}N^{-3/2}$ multiplied by the universal scaling function. Higher order terms are also determined to $O(1/N^4)$ in the $m = 0$ sector at $\widetilde{\Delta} = 1$. The latter series determines the first three terms of the KPZ to EW crossover scaling function for large argument where the scaling variable is $sN^{1/2}$. We then conclude with summary and discussion in Section VI.

## II. BETHE ANSATZ EQUATION

The asymmetric XXZ hamiltonian Eq. (1) is diagonalized by the Bethe ansatz. This is most conveniently seen from the fact that it is an anisotropic limit of the transfer matrix of the general asymmetric six-vertex model. This we show in Appendix A for self-containedness. Since Eq. (1) conserves $m$, we can consider each sector of fixed number of down-steps separately. We use the notation $Q$ to denote the number of down-steps (down-arrows in the language of the six vertex model) and let

$$q \equiv Q/N = (1 - m)/2. \tag{2}$$

The energy of Eq. (1) in the sector $Q$ is given by

$$E_N = \sum_{j=1}^{Q}\{\widetilde{\Delta} - \frac{1+s}{2} z_j^{-1} - \frac{1-s}{2} z_j\} \tag{3}$$

where the complex fugacities $\{z_j\}$ satisfy the Bethe ansatz equation,

$$z_i^N = (-1)^{Q-1} \prod_{j=1}^{Q} \frac{1 + e^{-4H} z_i z_j - 2\Delta e^{-2H} z_i}{1 + e^{-4H} z_i z_j - 2\Delta e^{-2H} z_j}. \tag{4}$$

The notation $z$ used in this section and Appendix A refers to the fugacity and should not be confused with the dynamic exponent. Different solutions of Eq. (4) give the energy of different eigenstates. Here, we have used the standard six-vertex model notations of [17] for $H$ and $\Delta$ which are related to $s$ and $\widetilde{\Delta}$ of Eqs. (1) and (3) by

$$\widetilde{\Delta} = \Delta/\cosh(2H) \tag{5a}$$

$$s = \tanh(2H). \tag{5b}$$

Since Eq. (1) with $s < 0$ is related to that with $s > 0$ by the simple particle-hole symmetry, we only consider the region $0 < s \leq 1$. We are interested in the region near $\widetilde{\Delta} = 1$ and parametrize $\Delta$ by

$$\Delta = \cosh \nu \tag{6}$$

($\nu > 0$). The fugacity variable $z$ is transformed to a more convenient variable $x$ defined through

$$z = \exp(2H - \nu)\frac{1-x}{1-\alpha x} \tag{7}$$

where

$$\alpha = \exp(-2\nu). \tag{8}$$

We note here that the five-vertex limit is achieved by letting $\nu$ and $H$ approach infinity ($\alpha \to 0$), keeping $2H - \nu$ fixed.

Using these notations, Eqs. (3) and (4) then take the simpler forms,

$$E_N = \frac{\sinh \nu}{\cosh 2H} \sum_{j=1}^{Q} \{-\frac{x_j}{1-x_j} + \frac{\alpha x_j}{1-\alpha x_j}\}, \tag{9}$$

and

$$\left(e^{2H-\nu}\frac{1-x_i}{1-\alpha x_i}\right)^N = (-1)^{Q-1}\prod_{j=1}^{Q}\frac{x_i - \alpha x_j}{x_j - \alpha x_i}, \tag{10}$$

respectively. Next we introduce, following de Vega and Woynarovitch [18], the function of complex variable

$$iZ_N(x) = (2H-\nu) + \log\frac{1-x}{x^q(1-\alpha x)} + \frac{1}{N}\sum_{j=1}^{Q}\{\log x_j - \log\frac{1-\alpha x_j/x}{1-\alpha x/x_j}\} \tag{11}$$

and its derivative

$$R_N(x) = ixZ_N'(x). \tag{12}$$

$Z_N(x)$ is called the phase function in this work and plays a central role below. It depends on $\{x_j\}$ , $H$, $\nu$ and $q$. Eq. (10) then can be written as

$$Z_N(x_j) = \frac{2\pi}{N}I_j \tag{13}$$

where $I_j$ are half-integers (integers) for $Q$ even (odd). The set $\{I_j\}$ specifies a particular energy level and is related to the momentum of the state by

$$P \equiv \frac{1}{i}\sum_{j=1}^{Q}\log z_j = \frac{2\pi}{N}\sum_{j=1}^{Q} I_j. \tag{14}$$

It is well established that the ground state is obtained if $\{I_j\}$ is chosen as

$$I_j = -\frac{Q+1}{2} + j \quad \text{for} \quad j = 1, \ldots, Q \tag{15}$$

while low lying excited levels are obtained by choosing sets of $\{I_j\}$ differing from that of the ground state in a few places near the end. For example, at $\widetilde{\Delta} = 1$, the first excited state is associated with the set

$$I_j = -\frac{Q+1}{2} + j \quad \text{for} \quad j = 1, \ldots, Q-1, \quad \text{and} \quad I_Q = \frac{Q+1}{2} \tag{16}$$

and has momentum $P = 2\pi/N$, while the first excited state in the $P = 0$ subspace is associated with the set

$$I_1 = -\frac{Q+1}{2}, \quad I_j = -\frac{Q+1}{2} + j \quad \text{for} \quad j = 2, \ldots, Q-1, \quad \text{and} \quad I_Q = \frac{Q+1}{2}. \tag{17}$$

In the thermodynamic limit $N \to \infty$, the roots $x_j$ for the ground state and its neighboring states form a continuous contour in the complex-$x$ plane. It is the inverse image, $Z_\infty^{-1}(\phi)$, of the segment of the real axis $-\pi q < \phi < \pi q$ of the complex $\phi$ plane. Here $Z_N^{-1}$ denotes the inverse function of $Z_N$. $R_\infty(x)/2\pi$ is then the density of roots along the contour. In particular when $\widetilde{\Delta} = 1$, the contour turns out to be a closed contour starting from $e^{\pi i} q/(1-q)$ and ending at

$$x_c^0 \equiv e^{-\pi i} q/(1-q) \tag{18}$$

enclosing the origin clockwise, as $\phi$ varies from $-\pi q$ to $\pi q$ [13,19]. In other words, we have

$$Z_\infty(x_c^0) = \pi q \quad \text{and} \quad Z_\infty(x_c^0 e^{2\pi i}) = -\pi q. \tag{19}$$

The explicit form of $Z_\infty$ for $\widetilde{\Delta} = 1$ is

$$i Z_\infty(x) = \log \frac{1-x}{x^q} + \log(q^q (1-q)^{1-q}). \tag{20}$$

This is obtained by integrating the result for $R_\infty(x)$ of [13] which is $R_\infty(x) = -q - x/(1-x)$ in our notation and by fixing the integration constant using Eq. (11). The branch cut for the log function is along the negative real axis. An actual form of the contour is shown in Fig. 1 for the case of $q = 1/2$. It is to be noted here that it has a cusp at $x = x_c^0$ due to the fact that $Z'_\infty(x_c^0) = 0$. Therefore, if one performs a power series expansion of Eq. (20) at $x_c^0$, the first order term vanishes and the result becomes

$$Z_\infty(x) = \pm \pi q - i \frac{(1-q)^3}{2q}(x - x_c^0)^2 + O((x - x_c^0)^3) \tag{21}$$

where the $+(-)$ sign is for $x$ below (above) the branch cut. Accordingly, $Z_\infty^{-1}(\phi)$ has the square-root type singularity at $\phi = \pm \pi q$ and takes the form

$$Z_\infty^{-1}(\pi q - \xi) = x_c^0 - i \sqrt{\frac{2q}{(1-q)^3}} \sqrt{i\xi} + O(\xi) \tag{22}$$

near the upper end of $\phi$ and

$$Z_\infty^{-1}(-\pi q + \xi) = x_c^0 + i \sqrt{\frac{2q}{(1-q)^3}} \sqrt{-i\xi} + O(\xi) \tag{23}$$

near the lower end of $\phi$. We shall see later that this square-root singularity is the origin of the singularity of free energy as a function of $\widetilde{\Delta}$ [13] and the unusual $1/N^{3/2}$ scaling of the mass gaps [5].

## III. SUMMATION FORMULA

In this section we assume that the phase function $Z_N(x)$ is known and derive a general expression of finite sums in terms of its expansion coefficients at a critical point $x_c$. The problem at hand is then to evaluate sums of the form (See Eqs. (9) and (11).)

$$\begin{aligned} S[f] &= \sum_{j=1}^{Q} f(x_j) \\ &= \sum_{j=1}^{Q} f(Z_N^{-1}(\frac{2\pi}{N} I_j)) \\ &= \sum_{j=1}^{Q} f(Z_N^{-1}(-\pi q + \frac{2\pi}{N}(j - \frac{1}{2}))) + \text{other terms} \end{aligned} \tag{24}$$

where '*other terms*' above denotes the terms which differ from that of the ground state; for example, for the first excited state characterized by Eq. (16),

$$\text{other terms} = f(Z_N^{-1}(\pi q + \frac{\pi}{N})) - f(Z_N^{-1}(\pi q - \frac{\pi}{N})), \tag{25}$$

while more terms are added and subtracted for other levels. In the following, we assume $S[f]$ is for the first excited state for the sake of notational simplicity and comment on other levels when appropriate.

To evaluate the finite sum in Eq. (24), we use the sum formula employed by Gwa and Spohn [5,20];

$$\sum_{j=1}^{Q} F(j) = \int_1^Q F(t)\,dt + \frac{1}{2}(F(Q) + F(1)) + 2\int_0^\infty \frac{\widetilde{F}(Q,t) - \widetilde{F}(1,t)}{e^{2\pi t} - 1}\,dt \tag{26}$$

where $\widetilde{F}(s,t) = (F(s+it) - F(s-it))/2i$. The finite size corrections of the sum are determined from properties of the summand near the end points of the sum. To produce the $1/N^{3/2}$ scaling correctly, the effect of the square-root singularity discussed above should be taken account. Thus, we first assume that $Z_N(x)$ has a vanishing derivative at a certain value of $x$ and define $x_c$ by the relation

$$Z_N'(x_c) = 0. \tag{27}$$

For large $N$, $x_c$ is expected to be close to $x_c^0$ but may pick up an imaginary part in which case we let the branch cut of log function pass through $x_c$. Next we assume that $Z_N'(x)$ is analytic at $x_c$. This assumption is based on the observation that $R_N(x) = -q - x/(1-x)$ exactly in the five-vertex limit and that expansion of Eq. (11) in the power series of $\alpha$ does not introduce extra singularities. Also define $\delta_\pm$ by the relation

$$Z_N(x_c) = \pi q - i\delta_+ \quad \text{and} \quad Z_N(x_c e^{2\pi i}) = -\pi q - i\delta_-. \tag{28}$$

Comparing Eq. (28) with Eq. (19), we also expect $\delta_\pm$ to be small for small $1 - \widetilde{\Delta}$ and large $N$. Presence of such terms is crucial in the expansion method we are describing below and has been anticipated from the solutions of the five-vertex model by a different method [21].

Definitions of $x_c$ and $\delta_\pm$ together with the assumption of analyticity of $Z_N'(x)$ at $x_c$ then allow us to put $Z_N^{-1}$ in the form

$$Z_N^{-1}(\pi q - \xi) = x_c + \sum_{m=1}^{\infty} a_m(-i\sqrt{\delta_+ + i\xi})^m \tag{29}$$

for the upper end and

$$Z_N^{-1}(-\pi q + \xi) = x_c + \sum_{m=1}^{\infty} a_m(i\sqrt{\delta_- - i\xi})^m \tag{30}$$

for the lower end of the sums in Eq. (24). Eqs. (29) and (30) implies that $Z_N(x)$ has the expansion

$$iZ_N(x) = \pm\pi q i + \delta_\pm + \frac{1}{a_1^2}(x - x_c)^2 - \frac{2a_2}{a_1^4}(x - x_c)^3 + \cdots. \tag{31}$$

$x_c$, $\delta_\pm$ and $a_m$ are to be determined self-consistently. They are in general complex but are defined in such a way that their leading term in the following perturbation expansion is real. Before proceeding further, we argue *a posteriori* that $\delta_+ = \delta_-$. The following steps of the derivation of $Z_N(x)$ show that it is of the form $iq\log x +$ (*parts analytic at* $x_c$) regardless of $\delta_+ = \delta_-$ or not. Then $Z_N(x_c) - Z_N(x_c e^{2\pi i}) = 2\pi q$ implying $\delta_+ = \delta_-$ from Eq. (28). Hence, we put

$$\delta_+ = \delta_- = \delta \tag{32}$$

from now on.

Then, the summand of Eq. (24) near the upper and lower ends can be written in the form

$$f(Z_N^{-1}(\pi q - \xi)) = f(x_c) + \sum_{m=1}^{\infty} A_m[f]\,(-i\sqrt{\delta + i\xi})^m, \tag{33}$$

and

$$f(Z_N^{-1}(-\pi q + \xi)) = f(x_c e^{2\pi i}) + \sum_{m=1}^{\infty} A_m[f]\,(i\sqrt{\delta - i\xi})^m, \tag{34}$$

respectively. In the above, the coefficient $A_m[f]$ is given by

$$A_m[f] = \sum_{k=1}^{m} \frac{b_{m,k}}{k!} f^{(k)}(x_c) \tag{35}$$

where $b_{m,k}$ $(1 \leq k \leq m)$ is related to $a_m$ through the relation

$$(\sum_{m=1}^{\infty} a_m \epsilon^m)^k = \sum_{m=k}^{\infty} b_{m,k} \epsilon^m, \tag{36}$$

and $f^{(k)}$ denotes the $k$-th derivative of $f(x)$. When Eq. (26) is applied to Eq. (24), the first term on the right hand side of Eq. (26) contributes

$$\begin{aligned} \frac{N}{2\pi} \int_{-\pi q+\pi/N}^{\pi q-\pi/N} f(Z_N^{-1}(\phi))\, d\phi = & -\frac{N}{2\pi} \oint f(x) Z_N'(x)\, dx \\ & + \frac{N}{2\pi} \left\{ \int_{\pi q - i\delta}^{\pi q-\pi/N} f(Z_N^{-1}(\phi))\, d\phi + \int_{-\pi q+\pi/N}^{-\pi q - i\delta} f(Z_N^{-1}(\phi))\, d\phi \right\} \end{aligned} \tag{37}$$

where $\oint$ denotes the integral along the closed contour in $x$-plane in the counterclockwise-sense. Using Eqs. (33) and (34) for the correction terms in Eq. (37) and the rest of the terms in Eq. (26), one can easily see that the $m$-th order terms in the series of Eqs. (33) and (34) contribute to order $N^{-m/2}$ in $S[f]$, provided the variable

$$y \equiv \frac{\delta N}{\pi} \tag{38}$$

is fixed to a finite value. To collect the $O(N^{-m/2})$ contributions, we are thus led to define a set of scaling functions $Y_m(y)$ as follows;

$$Y_m(y) = Y_m^0(y) + (-i\sqrt{y-i})^m - (-i\sqrt{y+i})^m, \tag{39}$$

where

$$Y_m^0(y) = \mathrm{Re}\left( \frac{m+2iy}{m+2}(-i\sqrt{y+i})^m + \frac{1}{i}\int_0^{\infty} \frac{(-i\sqrt{y+i+t})^m - (-i\sqrt{y+i-t})^m}{e^{\pi t}-1}\, dt \right) \tag{40}$$

is the contribution from Eq. (26) and the last two terms in Eq. (39) come from the two terms of Eq. (25). Useful properties of $Y_m$ and $Y_m^0$ are discussed in Appendix B. In terms of $Y_m$'s thus defined, Eq. (24) then can be written in the form of a series

$$S[f] = -\frac{N}{2\pi i} \oint f(x) R_N(x)\, \frac{dx}{x} + \frac{i}{2}(f(x_c) - f(x_c e^{2\pi i}) y + \sum_{m=1}^{\infty} A_m[f]\, Y_m(y)\, \varepsilon^m \tag{41}$$

where $\varepsilon$ is our perturbation expansion variable

$$\varepsilon \equiv \sqrt{\frac{\pi}{N}} \tag{42}$$

and $R_N$ is given by Eq. (12). The second term in Eq. (41) gives a contribution $\pi y$ for $f(x) = \log x$ but is zero for $f$ analytic at $x_c$.

## IV. GENERAL FORM OF $Z_N$ AND $E_N$

The sum formula Eq. (41) is derived pretending that $Z_N(x)$ is known. One then applies Eq. (41) to Eq. (11) which is the defining equation of $Z_N$ to determine it self-consistently. Here, $f = f_Z$ where

$$f_Z(x') = \log x' - \log \frac{1-\alpha x'/x}{1-\alpha x/x'}. \tag{43}$$

This then leads to an integral equation for $R_N(x)$ which is solved by Fourier method. To handle the contour integrations, we represent $f_Z$ as a Laurent series

$$f_Z(x') = \log x' + \sum_{n\neq 0} \frac{\alpha^{|n|}}{n} (\frac{x'}{x})^n. \tag{44}$$

This is justified since $0 \le \alpha < 1$ and integrations can be made keeping $|x'/x| = 1$. Using Eqs. (41) and (44) in Eq. (11) then gives us

$$iZ_N(x) = -q \log x + G(x) - \frac{1}{2\pi i} \oint f_Z(x')\, R_N(x')\, \frac{dx'}{x'} \tag{45}$$

where

$$\begin{aligned} G(x) &= (2H - \nu) + \log \frac{1-x}{1-\alpha x} + \delta \\ &- \frac{1}{\pi} \sum_{m=1}^{\infty} \sum_{k=1}^{m} b_{m,k} \{ \sum_n \frac{(-1)^k (n+1)(n+2)\cdots(n+k-1)}{k! x_c^k} \alpha^{|n|} (\frac{x}{x_c})^n \} Y_m(y)\, \varepsilon^{m+2}. \end{aligned} \tag{46}$$

Here and below $\sum_n$ stands for the sum over all integers $n$. Let us denote the Fourier mode of $G(x)$ by $G_n$;

$$G(x) = \sum_n G_n x^n. \tag{47}$$

Taking derivative of both sides of Eq. (45) with respect to $x$, one then obtains the integral equation for $R_N(x)$;

$$R_N(x) = -q + \sum_n n G_n x^n - \frac{1}{2\pi i} \oint f_R(x')\, R_N(x')\, \frac{dx'}{x'} \tag{48}$$

where

$$f_R(x') = -\sum_{n\neq 0} \alpha^{|n|} (\frac{x'}{x})^n. \tag{49}$$

The solution of Eq. (48) is

$$R_N(x) = -q + \sum_{n\neq 0} \frac{n G_n}{1-\alpha^{|n|}}\, x^n. \tag{50}$$

Putting this back into Eq. (45), one obtains

$$iZ_N(x) = \pi q i - q \log \frac{x}{x_c} + G_0 + \sum_{n\neq 0} \frac{G_n}{1-\alpha^{|n|}} (x^n - x_c^n). \tag{51}$$

Inserting explicit forms of $G_n$, we then finally obtain

$$\begin{aligned} iZ_N(x) &= \pi q i + (2H - \nu) + \delta - \frac{1}{\pi} \sum_{m=1}^{\infty} \sum_{k=1}^{m} b_{m,k} \frac{(-1)^k}{k x_c^k} Y_m(y)\, \varepsilon^{m+2} \\ &- q \log \frac{x}{x_c} + \log \frac{1-x}{1-x_c} + \frac{1}{\pi} \sum_{m=1}^{\infty} \sum_{k=1}^{m} b_{m,k} \{ g_k(x) - g_k(x_c) \}\, Y_m(y)\, \varepsilon^{m+2} \end{aligned} \tag{52}$$

with

$$g_k(x) = \frac{(-1)^{(k-1)}}{k! x_c^k} \sum_{n\neq 0} (n+1)(n+2)\cdots(n+k-1) \frac{\alpha^{|n|}}{1-\alpha^{|n|}} (\frac{x}{x_c})^n. \tag{53}$$

The series in Eq. (53) is convergent for $\alpha < |x/x_c| < 1/\alpha$. In deriving Eq. (52), we assume $|x| < 1$ which is valid for $q < 1/2$. However, final results can be extended to the region $q \ge 1/2$. $Z_N$ in the five-vertex limit ($\alpha \to 0$,

$2H-\nu=-\log\widetilde{\Delta}$) differs from that of the general six-vertex case only by the last double sum of Eq. (52) which is $O(\varepsilon^3)$.

To determine $\delta$ (i.e. $y$), $x_c$, and $a_m$'s, we use their defining equations in Eq. (52). Eq. (28) then gives

$$2H-\nu=\frac{1}{\pi}\sum_{m=1}^{\infty}\sum_{k=1}^{m}b_{m,k}\frac{(-1)^k}{kx_c^k}Y_m(y)\varepsilon^{m+2}, \tag{54a}$$

Next, comparing $iZ_N^{(k)}(x_c)$ derived from Eq. (52) with the coefficient of $(x-x_c)^k$ of Eq. (31), we obtain an infinite hierarchy of equations;

$$0=-\frac{q}{x_c}-\frac{1}{1-x_c}+\frac{1}{\pi}\sum_{m=1}^{\infty}\sum_{k=1}^{m}b_{m,k}g_k'(x_c)Y_m(y)\varepsilon^{m+2}, \tag{54b}$$

$$\frac{2}{a_1^2}=\frac{q}{x_c^2}-\frac{1}{(1-x_c)^2}+\frac{1}{\pi}\sum_{m=1}^{\infty}\sum_{k=1}^{m}b_{m,k}g_k''(x_c)Y_m(y)\varepsilon^{m+2}, \tag{54c}$$

$$\frac{-12a_2}{a_1^4}=\frac{-2q}{x_c^3}-\frac{2}{(1-x_c)^3}+\frac{1}{\pi}\sum_{m=1}^{\infty}\sum_{k=1}^{m}b_{m,k}g_k'''(x_c)Y_m(y)\varepsilon^{m+2}, \tag{54d}$$

etc. A merit of Eq. (54) is that it can be used to determine $y$, $x_c$ and the $a_m$'s order by order in $\varepsilon$.

Having determined, though formally, $Z_N(x)$, we next apply Eq. (41) to Eq. (9) using $f=f_E$ where

$$f_E=-\sum_{n=1}^{\infty}(1-\alpha^n)x^n. \tag{55}$$

If $R_N(x)$ is of the form of Eq. (50), then a simple application of the residue theorem gives

$$\frac{1}{2\pi i}\oint f_E(x)\,R_N(x)\,\frac{dx}{x}=\sum_{n=1}^{\infty}nG_{-n}. \tag{56}$$

This simple result is due to the particular form of $f_E$. Combining this with the explicit form of $f_E^{(k)}(x_c)$, we find

$$E_N=-\frac{\sinh\nu}{\cosh 2H}\sum_{m=1}^{\infty}\sum_{k=1}^{m}\frac{b_{m,k}}{(1-x_c)^{k+1}}\,Y_m(y)\,\varepsilon^m. \tag{57}$$

It is convenient to eliminate the first term in the double series of Eq. (57) using Eq. (54a) to make the series start with $O(\varepsilon^2)$ term. Thus we write $E_N$ as

$$E_N=\frac{\sinh\nu}{\cosh 2H}\left\{\frac{x_c}{(1-x_c)^2}(2H-\nu)N+T_N\right\} \tag{58a}$$

where

$$T_N=\sum_{m=2}^{\infty}\sum_{k=2}^{m}C_k[x_c]\,b_{m,k}Y_m(y)\varepsilon^m \tag{58b}$$

with

$$C_k[x_c]=\frac{(x_c-1)^{k-1}-kx_c^{k-1}}{k(1-x_c)^{k+1}x_c^{k-1}}. \tag{58c}$$

Eq. (58) together with (54) is the central result of this work. For energy levels other than the first excited one, one needs to replace $Y_m(y)$'s in Eqs. (58) and (54) by appropriate ones; e.g. by $Y_m^0(y)$ for the ground state energy. $Y_m$'s for the energy level of Eq. (17) will be discussed at the end of the next section and in Appendix B.

## V. LEADING ORDER BEHAVIORS OF THE MASS GAPS

We now find perturbative solutions to Eq. (54) and discuss their consequences. To the zeroth order in $\varepsilon$, $x_c$ and $a_m$ take the values $x_c^0$ and $a_m^0$, respectively, where $x_c^0$ is given by Eq. (18) and $a_m^0$'s are obtained by solving Eqs. (54c), (54d), etc. with $\varepsilon = 0$ or, more conveniently, by inverting the Taylor expansion of $Z_\infty(x)$ around $x_c^0$, Eq. (21). The first few $a_m^0$'s are

$$a_1^0 = \sqrt{\frac{2q}{(1-q)^3}}, \qquad a_2^0 = -\frac{2(1+q)}{3(1-q)^2} \quad \text{and} \quad a_3^0 = a_1^0 \frac{1+11q+q^2}{18q(1-q)}. \tag{59}$$

Inspection of Eqs. (54c), (54d), etc. shows that $a_m = a_m^0 + O(\varepsilon^3)$. Furthermore, the $O(\varepsilon^3)$ term of Eq. (54b) is proportional to $g_1'(x_c)$ which can be shown to be identically zero from its definition, Eq. (53). Thus, we have $x_c = x_c^0 + O(\varepsilon^4)$. Therefore, using $x_c = x_c^0$ and $a_m = a_m^0$ in Eq. (54a) and (57), we get correct expressions to $O(\varepsilon^5)$ and $O(\varepsilon^3)$, respectively. Also, since $2H - \nu = O(\varepsilon^3)$, and $\widetilde{\Delta} = 1 - s(2H-\nu) + O((2H-\nu)^2)$ from Eq. (5), we can put $2H - \nu = (1-\widetilde{\Delta})/s$ correct to order $\varepsilon^5$. Taking these considerations into account and using Eq. (36) , Eq. (54a) then becomes

$$1 - \widetilde{\Delta} = \frac{s}{\pi}\left\{\sqrt{\frac{2}{q(1-q)}}\, Y_1(y)\,\varepsilon^3 + \frac{1-2q}{3q(1-q)} Y_2 \varepsilon^4 + \sqrt{\frac{2}{q(1-q)}}\,\frac{1-q+q^2}{18q(1-q)}\, Y_3(y)\,\varepsilon^5 + O(\varepsilon^6)\right\} \tag{60}$$

which can be used to determine $y$ as a function of $(1-\widetilde{\Delta})$. This suggest that the appropriate scaling variable is

$$t = \sqrt{\frac{q(1-q)}{2\pi}}\left(\frac{1-\widetilde{\Delta}}{s}\right) N^{3/2}. \tag{61}$$

To the leading order in $\varepsilon$ , $y$ is related to $t$ by

$$t = Y_1(y). \tag{62}$$

Similarly, Eq. (58a) becomes

$$E_N = -q(1-q)(1-\widetilde{\Delta})N + s\{(1-2q)\,Y_2\,\varepsilon^2 + \frac{2}{3}\sqrt{2q(1-q)}\,Y_3(y)\,\varepsilon^3 + O(\varepsilon^4)\}. \tag{63}$$

which gives $E_N$ as a function of $\widetilde{\Delta}$, $s = \tanh 2H$ and $q$ to leading orders. Here, $Y_2 = 2i$ is related to the momentum of the level by Eq. (B7) and $y$ is a function of the scaling variable $t$ through Eq. (62).

The mass gap is the difference of $E_N$ from the ground state energy $E_N^0$. Since the level dependences of $E_N$ enter only through definitions of $Y_m(y)$, $E_N^0$ is obtained by replacing $Y_m$ by $Y_m^0$ in Eqs. (63) and (62). Thus we have

$$E_N - E_N^0 = s(1-2q)2\pi i/N + \frac{2}{3}\{Y_3(y) - Y_3^0(y')\}s\sqrt{2q(1-q)}\,(\frac{\pi}{N})^{3/2} + O(1/N^2) \tag{64}$$

where $y$ is the solution of Eq. (62), $y'$ is the solution of $t = Y_1^0(y')$, and $Y_2^0 = 0$ is used. We remark that $Y_3(y) - Y_3^0(y')$ is a function of $t$ only with no other parameters. Thus we define a universal scaling function as

$$\begin{aligned}\mathcal{F}(t) &= \frac{\sqrt{2}}{3}\pi^{3/2}\{Y_3(y) - Y_3^0(y')\} \\ &= \frac{\sqrt{2}}{3}\pi^{3/2}\{Y_3(Y_1^{-1}(t)) - Y_3^0(Y_1^{0\ -1}(t))\}\end{aligned} \tag{65}$$

to put Eq. (64) in the form

$$E_N - E_N^0 = 2\pi i s m/N + s\sqrt{1-m^2}\mathcal{F}(t)/N^{3/2} + O(1/N^2) \tag{66}$$

where we recall from Eq. (2) that $m = 1 - 2q$. $\mathcal{F}(t)$ is the universal scaling function describing the crossover from the KPZ $1/N^{3/2}$ scaling to the isotropic $1/N$ scaling as will be shown shortly. It is universal in the sense that the real part of the mass gap, scaled by the factor $s\sqrt{1-m^2}/N^{3/2}$ depends on $s$, $m$, $1-\widetilde{\Delta}$ and $N$ only through the combination given by Eq. (61) as $N \to \infty$. $\mathcal{F}(t)$ is shown in Fig. 2 for $0 \le t < 4$. It is not defined for $t < 0$. Higher order terms in Eq. (66) give the correction to scaling.

For the single-step model, we have $\widetilde{\Delta} = 1$, which implies $t = 0$. In this case, $E_N^0$ is exactly 0 since $\mathcal{H}$ generates a stochastic process. This is reflected in Eq. (65) by the fact that when $t \to 0^+$, $y' \to -\infty$ and $Y_3^0(y') \to 0^-$ as discussed in Appendix B. Thus we have $\mathcal{F}(0) = \sqrt{2}\pi^{3/2} Y_3(y_0)/3$ where $y_0$ is the solution of $Y_1(y_0) = 0$. Numerically, $\mathcal{F}(0) = 6.50918933794\ldots$. This is exactly the amplitude of the mass gap obtained by Gwa and Spohn [5] for $s = 1$ and $m = 0$. The dynamic exponent $z$ is then $3/2$ for any $s$ and $m$. Further implicatons of Eq. (66) are discussed in the next section.

When $1 - \widetilde{\Delta}$ is finite and positive, $t \to \infty$ for large $N$. In this case, the model is in the critical phase described by conformal field theory with central charge $c = 1$ and hence is expected to possess mass gaps which scale as $1/N$ [22]. To check this, we use the large $y$ behaviors of $Y_m(y)$ and $Y_m^0(y')$ discussed in Appendix B to obtain $Y_3(y)$ and $Y_3^0(y')$ as a function of $t$. The result is

$$Y_3^0(y') = -\frac{2}{5}\left(\frac{3t}{2}\right)^{5/3} - \frac{1}{6}\left(\frac{3t}{2}\right)^{1/3} + O(t^{-1}), \tag{67a}$$

$$Y_3(y) = -\frac{2}{5}\left(\frac{3t}{2}\right)^{5/3} + \frac{11}{6}\left(\frac{3t}{2}\right)^{1/3} + O(t^{-1}). \tag{67b}$$

From Eq. (67), we have

$$\mathcal{F}(t) = \frac{2\sqrt{2}\pi^{3/2}}{3}\left(\frac{3t}{2}\right)^{1/3} + O(t^{-1}) \tag{68}$$

as $t \to \infty$. The first terms in Eq. (67) contribute to the $O(N)$ bulk term in Eq. (63), while the second terms give the $O(1/N)$ correction terms. Using Eqs. (68), (61) and (2) in Eq. (66), we then have

$$E_N - E_N^0 = \frac{2\pi}{N}\left\{smi + \left(\frac{\pi}{18}s^2(1-m^2)^2(1-\widetilde{\Delta})\right)^{1/3}\right\}\{1 + O((1-\widetilde{\Delta})^{2/3})\} + O(1/N^2). \tag{69}$$

Also, using Eqs. (67), (61) and (2) in Eq. (63), we have the bulk energy

$$\lim_{N\to\infty} E_N^0/N = -\frac{1}{4}(1-m^2)(1-\widetilde{\Delta}) - \frac{1}{20}\left(\frac{3\pi}{2}\right)^{2/3} s^{-2/3}(1-m^2)^{4/3}(1-\widetilde{\Delta})^{5/3}\{1 + O((1-\widetilde{\Delta})^{2/3})\}. \tag{70}$$

The $O((1-\widetilde{\Delta})^{2/3})$ corrections in Eqs. (69) and (70) come from higher order terms of $\varepsilon$ not shown in Eq. (63). We note here that both the mass gap scaling and the ground state energy singularity in $1 - \widetilde{\Delta}$ is determined by the $\varepsilon^3 \sim N^{-z}$ term of Eq. (63). The $t^{5/3}$ behavior of $Y_3^0(y')$ and $Y_3(y)$ in Eq. (67) is necessary to produce $O(N)$ terms when combined with the $\varepsilon^3$ factor. Therefore the ground state energy singularity exponent $5/3$ is related to the dynamic exponent by $1 + 1/z$ as noted in [7] for the six vertex model. We also note that these fractional powers result from the fact that $\varepsilon \sim N^{-1/2}$ which in turn originates from the square-root singularity of $Z_\infty^{-1}(\phi)$ at $\phi = \pm\pi q$ as mentioned in Section II.

To the order discussed so far, the growth rate ($s$) dependence of the mass gap is rather trivial. Recently, Neergaard and den Nijs [6] have studied the mass gaps in the $q = 1/2$ (i.e. $m = 0$) sector numerically to discuss the crossover from the KPZ $1/N^{3/2}$ scaling to the EW $1/N^2$ scaling as $s \to 0$ for $\widetilde{\Delta} = 1$. To obtain nontrivial $s$-dependence, we need to include higher order corrections in Eq. (58). As discussed above, $x_c$ and $a_m$ differ from their zeroth order value by $O(\varepsilon^4)$ and $O(\varepsilon^3)$, respectively. We use these zeroth order values in the double sums of Eq. (54). This produces errors of $O(\varepsilon^7)$ in Eq. (54b) and $O(\varepsilon^6)$ in Eq. (54c), (54d), etc. We then solve for $x_c$ and $a_m$ perturbatively from the approximated Eq. (54) to obtain $x_c$ correct to $O(\varepsilon^6)$ and $a_m$ to $O(\varepsilon^5)$. Using those values in Eq. (58b) would give correct result to $O(\varepsilon^7)$. However, there occurs an extra simplification for $q = 1/2$. The first term of Eq. (58b) is proportional to $C_2[x_c]a_1^2$ which becomes simply $1 - 2q$ at the zeroth order. Thus the knowledge of $a_1$ to $O(\varepsilon^5)$ is sufficient to determine it to $O(\varepsilon^6)$, resulting in $T_N$ correct to $O(\varepsilon^8)$. Symbolic manipulation softwares are used for the algebra. The result for $T_N$ at $q = 1/2$ is

$$\begin{aligned} T_N = {} & \frac{\sqrt{2}}{3}Y_3(y)\varepsilon^3 - \frac{\sqrt{2}}{10}Y_5(y)\varepsilon^5 + \frac{8}{\pi}\beta_2(\alpha)(Y_1(y)Y_3(y) - Y_2^2)\varepsilon^6 + \frac{5}{84\sqrt{2}}Y_7(y)\varepsilon^7 \\ & + \frac{1}{\pi}\{\frac{4}{3}(\beta_2(\alpha) + 8\beta_4(\alpha))Y_3(y)^2 - 16\beta_4(\alpha)Y_2Y_4(y)\}\varepsilon^8 + O(\varepsilon^9) \end{aligned} \tag{71}$$

where $\beta_k(\alpha)$ ($k$ even) are functions of $\nu$ defined by

$$\beta_k(\alpha) = \sum_n \frac{n^k \alpha^{|n|}}{1-\alpha^{|n|}} \tag{72}$$

(See Eq. (8).) and $y$ is to be determined from Eq. (54a). When $\widetilde{\Delta}=1$, the equation for $y$ to the necessary order for $q=1/2$ becomes

$$0 = Y_1(y) + \frac{1}{6}Y_3(y)\varepsilon^2 + \frac{1}{120}Y_5(y)\varepsilon^4 + O(\varepsilon^5). \tag{73}$$

The solution of Eq. (73) can be obtained in the form $y = y_0 + y_1\varepsilon^2 + y_2\varepsilon^4 \cdots$ . Using it in Eq. (71) and using the relation $E_N = sT_N$ for $\widetilde{\Delta} = 1$ , we then finally obtain

$$E_N/s = \frac{\sqrt{2}}{3}Y_3(y_0)\varepsilon^3 - \frac{\sqrt{2}}{10}Y_5(y_0)\varepsilon^5 - \frac{8}{\pi}\beta_2(\alpha)Y_2^2\varepsilon^6 + \left\{\frac{5}{84\sqrt{2}}Y_7(y_0) + \frac{7\sqrt{2}}{72}\frac{Y_3(y_0)^2}{Y_{-1}(y_0)}\right\}\varepsilon^7$$
$$+\frac{1}{\pi}\{\frac{32}{3}Y_3(y_0)^2 - 16Y_2Y_4(y_0)\}\beta_4(\alpha)\varepsilon^8 + O(\varepsilon^9). \tag{74}$$

In deriving Eq. (74), we have used Eq. (B1) repeatedly. That $O(\varepsilon^2)$ and $O(\varepsilon^4)$ terms are missing is a simplifying feature of the $q=1/2$ sector. This is valid for $\widetilde{\Delta}=1$ only and $\beta_k$'s are now functions of $s$ since

$$\alpha = \exp(-4H) = \frac{1-s}{1+s} \tag{75}$$

when $\widetilde{\Delta}=1$. Thus the mass gap for $\widetilde{\Delta}=1$ and $q=1/2$ is of the form

$$E_N = \frac{s}{N^{3/2}}\left\{A_0 + \frac{A_1}{N} + \frac{A_2}{N^2}\right\} + s\beta_2(\alpha)\frac{B_0}{N^3} + s\beta_4(\alpha)\frac{B_1}{N^4} + O(N^{-9/2}). \tag{76}$$

where $A_j$ and $B_j$ are constants which are found numerically, except $B_0$. $A_0$ is nothing but $\mathcal{F}(0)$ introduced above. Numerical values of the coefficients are as follows;

$$A_0 = 6.50918933794\ldots, \qquad A_1 = 10.8607132611\ldots, \qquad A_2 = -10.3825789166\ldots,$$
$$B_0 = 32\pi^2, \qquad B_1 = -2407.62590782\ldots.$$

This expression for energy is in complete agreement with existing numerical values of $E_N$ obtained from diagonalization of Eq. (1) for small $N$'s ($N \leq 18$) in [6]. For example, $A_1/(A_0\pi)$ is estimated in [6] be to $0.531\pm0.002$ while it is $0.531106443911\ldots$ in our result. Moreover, the numerical sequence of $(E_N N^{3/2}/s - A_0 - A_1/N)N^2$ for $N$=6,8,...,18, obtained using the $s=1$ data of [6] together with the exact values of $A_0$ and $A_1$ given above extrapolates to around $-10.35 \sim -10.40$ consistent with the exact value of $A_2$. A similar procedure using the $s=0.999$ data could produce $B_0$ but with less precision [23].

It is of interest to consider the KPZ to EW crossover scaling function near $s=0$. When $s \to 0$, it can be shown that $\beta_k$ defined in Eq. (72) and (75) behaves as

$$\beta_k(\alpha) \to 2k!\zeta(k+1)(1-\alpha)^{-(k+1)} \sim k!2^{-k}\zeta(k+1)s^{-(k+1)} \tag{77}$$

where $\zeta$ is the Riemann zeta function. One notes from this that $\beta_2(\alpha)/N^3$ and $\beta_4(\alpha)/N^4$ in Eq. (76) are both $O(1/N^{3/2})$ provided $s\sqrt{N}$ is kept constant as $s \to 0$. Therefore the natural crossover scaling variable is $u = s\sqrt{N}$ and the mass gap takes the scaling form

$$E_N = \mathcal{G}(s\sqrt{N})/N^2 \tag{78}$$

as $s \to 0$ with $s\sqrt{N}$ fixed. Eq. (76) then gives the first three terms of the large argument expansion of the crossover scaling function $\mathcal{G}(u)$ as

$$\mathcal{G}(u) = A_0|u| + \frac{B_0'}{u^2} + \frac{B_1'}{u^4} + \cdots \tag{79}$$

where $B'_0 = \zeta(3)B_0/2$ and $B'_1 = 3\zeta(5)B_1/2$ . At the EW point ($s = 0$), it is known that $E_N = 2\pi^2/N^2$ exactly [6] and hence we have $\mathcal{G}(0) = 2\pi^2$. The scaling form Eq. (78) has been anticipated in [6]. However, the actual form of $\mathcal{G}$ proposed in [6] is of the form $2\pi^2\sqrt{1+u^2/\pi^2} = 2\pi|u| + \pi^3/|u| + \cdots$ which we find is valid only approximately.

So far, only the first excited state characterized by the set Eq. (16) is discussed. However, the above results are readily applied to other energy levels provided the scaling functions $Y_m(y)$ are substituted by appropriate ones. For example, to consider the momentum 0 level mentioned in Eq. (17), the $Y_m$ need to be changed to

$$Y_m(y) \to Y_m^0(y) + (-i\sqrt{y-i})^m + (i\sqrt{y+i})^m - (-i\sqrt{y+i})^m - (i\sqrt{y-i})^m \tag{80}$$

where $Y_m^0(y)$ are still given by Eq. (40) and the remaining 4 terms comes from two excitations and two holes in Eq. (17). Here, $Y_2 = 0$ from Eq. (B7). For $q = 1/2$, only the numerical values of the $Y_m(y_0)$ in Eq. (74) change. See Appendix B for the value of $y_0$. The coefficients of Eq. (76) are then found to be

$$A_0 = 16.0176269046\ldots, \qquad A_1 = 44.0512971116\ldots, \qquad A_2 = -7.2545175082\ldots,$$
$$B_0 = 0, \qquad B_1 = 12315.0898394\ldots.$$

In [6], $A_0/\pi^{3/2}$ for this level is estimated to be 2.87±0.01 which is consistent with our exact value 2.87655951907.... However, their conjectured exact value of 2.8633717... is off by 0.5%. Note that the non-trivial $s$ dependence appears only at $O(1/N^4)$. Other higher levels can be treated similarly.

## VI. SUMMARY AND DISCUSSION

In this work, we have developed a perturbative scheme which allows one to calculate the finite size corrections of the low lying energies of the asymmetric XXZ hamiltonian Eq. (1) near the stochastic line and obtained their leading order behaviors for arbitrary $s$ and arbitrary $m$. Formally, the energy is given by Eq. (58) together with Eq. (54) which determines perturbatively $y$, $x_c$ and $a_m$'s entering Eq. (58). We remind the reader that $q$ is trivially related to the substrate slope by Eq. (2) and other notations in Eq. (58) are defined in Section III.

The mass gap of the first excited level is found to take the scaling form given by Eq. (66) where the scaling variable $t$ is given by Eq. (61) and the universal scaling function $\mathcal{F}(t)$ is defined in Eq. (65) and shown in Fig. 2. $\mathcal{F}(t)$ behaves as $\sim t^{1/3}$ for $t \to \infty$ and $\mathcal{F}(0)$=constant. The result can also be applied to other levels. Thus if we use superscript $\ell$ to distinguish various low lying energy levels, we then have the scaling form for the mass gap as

$$E_N^\ell - E_N^0 = ismP_\ell + s\sqrt{1-m^2}\mathcal{F}_\ell(t)N^{-3/2} + O(N^{-2}) \tag{81}$$

with $P_\ell$=$(2\pi/N)\times$integer the momentum of the level $\ell$, $\mathcal{F}_\ell$ the level dependent scaling function. $\mathcal{F}_\ell(t)$ is also defined by Eq. (65) but with appropriate $Y_m$'s, e.g. Eq. (80) for the lowest level with $P\ell = 0$. This is the first main result of this paper.

When $\tilde{\Delta} \lesssim 1$, $t \to \infty$ as $N \to \infty$ and the mass gaps scale with $1/N$ as shown in Eq. (69) for the first excited level. In this case, the continuum theory of Eq. (1) is described by the central charge $c = 1$ conformal field theory. The first excited level characterized by Eq. (16) is the first descendant of the identity operator in the sense of conformal field theory. Thus the mass gap is expected to behave as $2\pi\tilde{\zeta}/N$, $\tilde{\zeta}$ being the complex anisotropy factor which is proportional to the modular ratio of the torus on which corresponding field theory is defined [24]. Eq. (69) then determines $\tilde{\zeta}$ for $\tilde{\Delta}$ close to 1, with the imaginary part accounting for the deformation of the square lattice to a parallelogram. The ground state energy itself shows a singularity of the type $(1-\tilde{\Delta})^{1+1/z}$ with $z = 3/2$. The crossover from the conformally invariant region $\tilde{\Delta} < 1$ to the KPZ line is described by the scaling functions $\mathcal{F}_\ell(t)$ as functions of the scaling variable $t$.

For the single-step growth model, $t = 0$, $E_N^0 = 0$ and $\mathcal{F}_\ell(t)$ becomes a universal constant $\mathcal{F}_\ell(0)$. A quantity of interest here is the stationary slope–slope correlation function given by [5]

$$G(r,\tau) = <0|\sigma_0^z e^{-|\tau|\mathcal{H}}\sigma_r^z|0> - m^2 \tag{82}$$

where $|0>$ is the ground state in sector $m$, $\tau$ stands for time and $G$ should not be confused with the notation used in Section IV. Our result then implies that the correlation function takes the form, after taking the spectral decomposition and using Eq. (81),

$$G(r,\tau) = \sum_\ell <0|\sigma_0^z|\ell><\ell|\sigma_0^z|0> \exp\{-iP_\ell(r+sm\tau) - s\sqrt{1-m^2}\mathcal{F}_\ell(0)|\tau|/N^{3/2}\} \tag{83}$$

as $\tau \to \infty$, $N \to \infty$ with $\tau/N^{3/2}$ fixed. Here, the sum is over excited levels whose mass gap scales as $1/N^{3/2}$, $|\ell>$ stands for excited state and the relation $\sigma_r^z = \exp(-i\mathcal{P}r)\sigma_0^z \exp(i\mathcal{P}r)$, $\mathcal{P}$ being the momentum operator whose eigenvalue is $P_\ell$, has been used. Therefore decay of the correlation is governed by a characteristic time which scales as $N^z$ with $z = 3/2$. This confirms that the dynamic exponent $z$ is $3/2$. It also confirms the universality with respect to the growth rate $(0 < s < 1)$ and the substrate slope $(-1 < m < 1)$. This is an assuring feature for the KPZ class growth models in view of the fact that, in the asymmetric exclusion process with reflecting boundary condition, the mass gaps remain finite even in the bulk limit [14]. The discrete Fourier transform of Eq. (83) gives the structure factor of finite size system in the scaling form

$$S(k,\tau) = e^{-iksm\tau}\mathcal{S}(s\sqrt{1-m^2}|\tau|k^{3/2}, kN) \tag{84}$$

for small $k$ as expected from other theoretical grounds [5]. The phase factor $e^{-iksm\tau}$ in Eq. (84) is due to the imaginary term of the mass gaps and is related to the steady state current as discussed in [15] and [5]. Presence of such a phase factor is also consistent with the invariance of the KPZ equation under the Galilean transformation which tilts the substrate by an infinitesimal $m$ and changes the spatial coordinate by $r \to r - sm\tau$ [2,25]. Thus, the modulating factor in Eq. (83) disappears along the ray $r + sm\tau = 0$ in space–time and the imaginary part of Eq. (81) offsets this shift in space coordinate.

Henkel and Schütz [15] showed that the mass gaps of Eq. (1) scale as $1/N^2$ for $q \sim O(1/N)$, i.e. for finite number of down steps instead of finite density of down steps. It is consistent with Eq. (81) due to the factor $\sqrt{1-m^2} = \sqrt{4q(1-q)}$. Higher order terms in Eq. (81) also contribute to $O(1/N^2)$ when $q \sim O(1/N)$. Even if the $1/N^2$ scaling for $q \sim O(1/N)$ is the same as that for the $s = 0$ EW point, their physical origins are different. The latter comes from the fact that correlation functions such as Eq. (82) satisfy the discrete diffusion equation exactly for all $m$ while the former originates from the non–relativistic free fermion dispersion relation near the band edge [3,5,6]. When $\tilde{\Delta} < 1$, the limits $q \to 0, 1$ are where the Pokrovsky–Talapov (PT) type commensurate–incommensurate transitions occur in the context of the domain wall picture as discussed in detail in [9] for the case of $s = 1$. The $1/N^2$ scaling of the mass gap near $q = 0$ is a signature of the PT transition in this case.

The second main result of this paper addresses the KPZ to EW crossover scaling function. The mass gaps Eq. (81) depend linearly with the growth rate $s$ to the leading orders whereas, when $s \to 0$ along the stochastic line, they are expected to cross over from the KPZ scaling to the EW scaling. Such non-trivial $s$-dependence comes from higher order terms in the perturbation series. Thus we have pushed the perturbative calculation to $O(1/N^4)$ for $m = 0$ (See Eqs. (74) and (76)). From the result, the KPZ to EW crossover scaling variable is found to be $u = s\sqrt{N} = sN^{2-z}$ and the mass gaps are shown to take the scaling form

$$E_N^\ell = \mathcal{G}_\ell(sN^{2-z})/N^2 \tag{85}$$

for $m = 0$ where the KPZ to EW crossover scaling function $\mathcal{G}_\ell(u)$ behaves as $\sim u$ for $u \to \infty$ and $\mathcal{G}_\ell(0)$=constant. The perturbative expansion scheme of this work allows one to obtain the expansion of the scaling function $\mathcal{G}_\ell(u)$ in powers of $1/u$. Our $O(1/N^4)$ result determines the first three terms in this expansion. The result is shown in Eq. (79) for the first excited level. For other levels, only the numerical values of the coefficients in the series change. Derivation of the exact form of $\mathcal{G}_\ell(u)$ is an open problem. Also, to what extent the results of this work remain valid in other KPZ class models as universal features is an interesting open question.

## ACKNOWLEDGMENTS

This work is supported by Ministry of Education, Republic of Korea, by Korea Science and Engineering Foundation through the Center for Theoretical Physics, Seoul National University, and also by NSF grant DMR-9205125. The author thanks Prof. M. den Nijs for motivating this work, helpful discussions and critical reading of the manuscript. He also thanks Profs. J. M. Kim and G. Schütz for valuable discussions.

## APPENDIX A: RELATION TO ASYMMETRIC SIX VERTEX MODEL

In Appendix A, we briefly review, for the sake of self-containment, relevant results of the asymmetric six-vertex model and derive the asymmetric XXZ hamiltonian as an anisotropic limit of it. Let $w_i$, $i$=1, ..., 6, denote the six Boltzmann weights of the six-vertex model in the standard labeling [17,26] . We denote by $\mathcal{T}$ the row-to-row transfer matrix of the general six-vertex model on $N \times N$ square lattice under periodic boundary conditions. The number of down-arrows, $Q$, in each row is conserved and there are four independent parameters parametrizing $w_i$'s. The eigenvalue $\Lambda$ of $\mathcal{T}$ in the sector of $Q$ down-arrows is given by

$$\Lambda = w_1^{N-Q} \prod_{j=1}^{Q} (w_3 + \frac{w_5 w_6 z_j}{w_1 - w_4 z_j}) + w_4^{N-Q} \prod_{j=1}^{Q} (w_2 - \frac{w_5 w_6}{w_1 - w_4 z_j}). \tag{A1}$$

The sets of fugacities $\{z_j\}$ are given by the Bethe ansatz equation Eq. (4) where in the present case,

$$H = \frac{1}{4} \log(w_1 w_3 / w_2 w_4) \tag{A2}$$

is the horizontal field and

$$\Delta = \frac{w_1 w_2 + w_3 w_4 - w_5 w_6}{2(w_1 w_2 w_3 w_4)^{1/2}} \tag{A3}$$

is the standard interaction parameter. It is also convenient to define

$$\tilde{\Delta} = \frac{\Delta}{\cosh(2H)} = \frac{w_1 w_2 + w_3 w_4 - w_5 w_6}{w_1 w_3 + w_2 w_4} \tag{A4}$$

and

$$s = \tanh(2H) = \frac{w_1 w_3 - w_2 w_4}{w_1 w_3 + w_2 w_4} \tag{A5}$$

for later uses. Note that $z_j$ in Eq. (A1) is the inverse of that defined in [17] and that $m = 1 - 2Q/N = 1 - 2q$ of this work corresponds to $y$ in standard six vertex notation. Eigenvectors of $\mathcal{T}$ depend only on $H$ and $\Delta$ and hence $\mathcal{T}$ form a two-parameter family of commuting matrices; i.e. $\mathcal{T}$'s with different values of $w_i$'s commute with each other as long as they have the same values of $H$ and $\Delta$. This is a direct consequence of the fact that the Yang-Baxter equation has a solution under the condition of constant $\Delta$ and $H$ [26]. When $w_i$'s assume the limiting values $w_1^0 = w_2^0 = w_5^0 = w_6^0 = 1$ and $w_3^0 = w_4^0 = 0$, $\mathcal{T}$ becomes the shift operator. We call this the anisotropic limit. Suppose $w_i$ are parametrized by $u$ and $v$ in addition to $H$ and $\Delta$ in such a way that the limit $u = v = 0$ corresponds to the anisotropic limit. An example of such parametrization for $\Delta > 1$ is

$$\begin{aligned} w_1 &= \exp(v) \sinh(u + \nu)/\sinh(\nu) \\ w_2 &= \exp(-v) \sinh(u + \nu)/\sinh(\nu) \\ w_3 &= \exp(2H - v) \sinh(u)/\sinh(\nu) \\ w_4 &= \exp(-2H + v) \sinh(u)/\sinh(\nu) \\ w_5 &= w_6 = 1 \end{aligned}$$

with $\Delta = \cosh(\nu)$ but actual form of parametrization is not important here.

The quantum chain hamiltonians commuting with $\mathcal{T}$ are obtained by taking derivatives of $-\log \mathcal{T}$ with respect to $u$ or $v$ at $u = v = 0$. A straightforward calculation yields [26]

$$\begin{aligned} \mathcal{H}' &= - \left. \frac{\partial \log \mathcal{T}}{\partial u} \right|_{u=v=0} \\ &= \sum_{i=1}^{N} \{ \frac{w_1' + w_2' - w_5' - w_6'}{4} (1 - \sigma_i^z \sigma_{i+1}^z) - w_3' \sigma_i^+ \sigma_{i+1}^- - w_4' \sigma_i^- \sigma_{i+1}^+ \} - w_1'(N - Q) - w_2' Q \end{aligned} \tag{A6}$$

for the sector Q, where $\sigma_i^\alpha$ are the Pauli spin operators and $w_i'$ stands for the derivative of $w_i$ with respect to $u$ at $u = v = 0$. Corresponding eigenvalue $E'$ of $\mathcal{H}'$ is, from Eq. (A1),

$$\begin{aligned} E' &= - \left. \frac{\partial \log \Lambda}{\partial u} \right|_{u=v=0} \\ &= \sum_{j=1}^{Q} \{ (w_1' + w_2' - w_5' - w_6') - w_3' z_j^{-1} - w_4' z_j \} - w_1'(N - Q) - w_2' Q. \end{aligned} \tag{A7}$$

For $\mathcal{H}'$ to commute with $\mathcal{T}$, the derivatives should be taken with $H$ and $\Delta$ fixed. This introduces two constraints;

$$w'_1 + w'_2 - w'_5 - w'_6 = \tilde{\Delta}(w'_3 + w'_4) \tag{A8a}$$

$$w'_3 = \exp(4H)w'_4. \tag{A8b}$$

Using Eq. (A8) in Eqs. (A6) and neglecting the constant terms, one then easily sees that $\mathcal{H}'$ becomes the asymmetric XXZ hamiltonian Eq. (1) up to a normalization factor $(w'_3 + w'_4)$. Therefore the energy of Eq. (1) is obtained from Eqs. (A7) and (A8) up to the same normalization factor and the constants, which is exactly Eq. (3).

## APPENDIX B: PROPERTIES OF $Y_M$

In Appendix B, we discuss some properties of $Y_m^0(y)$ and $Y_m(y)$ defined in Eqs. (40) and (39), respectively. Here, $Y_m^0$ refers to the scaling function associated with the ground state, while $Y_m$ refers to that with the first excited level.

The most useful property is the relation

$$Y'_m(y) = -\frac{m}{2} Y_{m-2}(y), \tag{B1}$$

which also holds for other $Y_m$'s associated with other levels including $Y_m^0$.

First we consider the case of $m$ odd. $Y_m^0$ and $Y_m$ are real for real $y$. As $y \to \infty$, their asymptotic forms are

$$Y_m^0 \to (-1)^{(m-1)/2} y^{(m+2)/2} \left( \frac{2}{m+2} + \frac{m}{12y^2} + \cdots \right), \tag{B2}$$

and

$$Y_m \to (-1)^{(m-1)/2} y^{(m+2)/2} \left( \frac{2}{m+2} - \frac{11m}{12y^2} + \cdots \right), \tag{B3}$$

respectively. To see how $Y_m^0(y)$ behaves as $y \to -\infty$, we expand Eq. (40) as a series in $-1/y$ paying attention to the branch cut of the square root. After a little algebra, we obtain

$$Y_m^0(y) = \sum_{k \geq 2, \mathrm{even}} (-1)^{k/2} \frac{m(m-2)\cdots(m-2k+2)}{2^k k!} \left( \frac{k}{k+1} - J_k \right) (-\frac{1}{y})^{k-m/2} \tag{B4}$$

where

$$J_k = \frac{1}{i} \int_0^\infty \frac{(1+it)^k - (1-it)^k}{e^{\pi t} - 1} dt. \tag{B5}$$

To calculate the integral $J_k$ above, we apply the sum formula Eq. (26) to the trivial sum $\sum_{j=0}^{1} (j - 1/2)^k$ and find that $J_k = k/(k+1)$ for $k$ even. Thus all coefficients of the series Eq. (B4) are identically zero. This suggests that $Y_m^0$ has an essential singularity at $y = -\infty$ and approaches to zero exponentially as $y \to -\infty$. For $Y_m(y)$ we have

$$Y_m(y) \to -2(-y)^{m/2} \{1 - \frac{m(m-2)}{8y^2} + \cdots\} \tag{B6}$$

as $y \to -\infty$. Numerically, we find that $Y_1^0$ and $-Y_3^0$ both are positive monotonic increasing function while $Y_1(y)$ is a monotonic increasing real function passing through 0 at $y_0 = 1.119066880828\ldots$. Due to the relation Eq. (B1), $y_0$ is the position of the maximum of $Y_3(y)$.

For $m$ even, one can show that $Y_m^0 = 0$ for all $m$ using the value of the integral $J_k$ while $Y_m$ is a polynomial in $y$ of order $(m-2)/2$. Specifically, $Y_2 = 2i$, $Y_4 = -4iy$, $Y_6 = i(6y^2 - 2)$, etc. and $Y_m \sim im(-y)^{(m-2)/2}$ for $y$ large.

For other energy levels than the first excited state we are considering here, the last terms of Eq. (39) are different. Thus, the numerical value of $y_0$ where $Y_1$ vanishes is different from level to level. For example, $Y_1$ defined in Eq. (80) has a zero at $y_0 = 1.65873919064\ldots$ However, one can show in general that $Y_2$ is related to the total momentum $P$ of Eq. (14) by

$$Y_2 = \frac{iN}{\pi} P. \tag{B7}$$

---

* Permanent address. E-mail: dkim@phyb.snu.ac.kr.

[1] See for example F. Family and T. Vicsek, *Dynamics of Fractal Surfaces* (World Scientific, Singapore, 1991).
[2] M. Kardar, G. Parisi and Y.-C. Zhang, Phys. Rev. Lett. **56**, 889 (1986).
[3] M. Plischke, Z. Rácz and D. Liu, Phys. Rev. B **35**, 3485 (1987), see also P. Meakin, P. Ramanlal, L. M. Sander and R. C. Ball, Phys. Rev. A **34**, 5091 (1986) for a special case of the model.
[4] T. M. Liggett, *Interacting Particle Systems* ( Springer-Verlag, Berlin, 1985), H. van Beijeren, R. Kutner and H. Spohn, Phys. Rev. Lett. **54**, 2026 (1985).
[5] L.-H. Gwa and H. Spohn, Phys. Rev. A **46**, 844 (1992), Phys. Rev. Lett. **68**, 725 (1992).
[6] J. Neergaard and M. den Nijs , Phys. Rev. Lett. **74** , 730 (1995).
[7] J. D. Shore and D. J. Bukman, Phys. Rev. Lett. **72**, 604 (1994).
[8] I. M. Nolden, J. Stat. Phys. **67**, 155 (1992).
[9] J. D. Noh and D. Kim, Phys. Rev. E **49**, 1943 (1994).
[10] J. L. Cardy, in *Phase Transitions and Critical Phenomena*, edited by C. Domb and J. L. Lebowitz, (Academic, London, 1987) Vol. 11.
[11] The mass gaps are in general complex and their scaling behavior refers to the $N$ dependence of their real part.
[12] See also D. Dhar, Phase Transitions **9**, 51 (1987).
[13] D. J. Bukman and J. D. Shore, J. Stat. Phys. **78**, 1277 (1995).
[14] S. Sandow and G. Schütz, Europhys. Lett. **26**, 7 (1994). G. M. Schütz, J. Stat. Phys. to appear.
[15] M. Henkel and G. Schütz, Physica A **206**, 187 (1994).
[16] H. C. Fogedby, A. B. Eriksson and L. V. Mikheev, preprint (cond-mat/9503028).
[17] E. H. Lieb and F. Y. Wu, in *Phase Transitions and Critical Phenomena*, edited by C. Domb and M. S. Green, (Academic, London, 1972) Vol. 1.
[18] H. J. de Vega and F. Woynarovitch, Nucl. Phys. B **251**, 439 (1985).
[19] $x$ in this work is related to $u$ of Ref. [13] by $x = \exp(-iu + b + \nu)$. Since $Z_\infty$ is related to the function $f(p^0)$ of Ref. [8] by an opposite sign, the sense of the contour of Fig. 1 is clockwise.
[20] J. Dieudonné, *Infinitesimal Calculus* (Hermann, Paris 1971) p.285.
[21] D. Kim, unpublished.
[22] J. D. Noh and D. Kim, unpublished.
[23] M. den Nijs, private communication.
[24] J. L. Cardy, Nucl. Phys. **B 270**, 186 (1986); C. Itzykson and J.-B. Zuber, Nucl. Phys. **B 275**, 580 (1986).
[25] The growth parameter $s$ of this work corresponds to $-\lambda$ of Ref. [2].
[26] R. J. Baxter, *Exactly Solvable Models in Statistical Mechanics* (Academic, London, 1982).

FIG. 1. Locus of $x = Z_\infty^{-1}(\phi)$ for $-\pi q < \phi \leq \pi q$ in the complex $x$ plane for the case of $q = 1/2$.

FIG. 2. The scaling function $\mathcal{F}(t)$ of the first excited energy level.